\documentclass[12pt]{iopart}
\usepackage{amssymb}

\usepackage{cite}
\usepackage{graphicx}

\begin{document}

\title{The confined hydrogen atom with a moving nucleus}
\author{Francisco M Fern\'andez}

\address{INIFTA (UNLP, CCT La Plata-CONICET), Divisi\'on Qu\'imica Te\'orica,
Blvd. 113 S/N,  Sucursal 4, Casilla de Correo 16, 1900 La Plata,
Argentina}\ead{fernande@quimica.unlp.edu.ar}

\maketitle

\begin{abstract}
We study the hydrogen atom confined to a spherical box with impenetrable
walls but, unlike earlier pedagogical articles on the subject, we assume
that the nucleus also moves. We obtain the ground--state energy
approximately by means of first--order perturbation theory and by a
more accurate variational approach. We show that it
is greater than the one for the case in which the nucleus is clamped at the
center of the box. Present approach
resembles the well--known treatment of the helium atom with clamped nucleus.
\end{abstract}

\section{Introduction}

\label{sec:intro}

Confined quantum--mechanical models have proved to be suitable first
approximations for estimating the effect of pressure on the spectral lines
of atoms and molecules or the effect of their neighbours in condense media.
Several such models have been proposed for pedagogical purposes in
introductory courses on quantum mechanics\cite
{R69,G75,AILZ81,E81b,E81a,L83,Y86,MC88,W89,BBM90,B91,MC91,T91,LM93,MRU95,RRML00,DC00,ACCG01,GB03,GRD06}%
. Among them we mention the quantum bouncer\cite{G75,AILZ81,L83}, the
harmonic oscillator\cite{MC88,T91,MC91,RRML00,ACCG01,GRD06} and the hydrogen
atom\cite{Y86,W89,B91,MC91,LM93,MRU95,DC00,GB03}. Such models have also been
useful for the discussion of semiclassical approaches\cite{L83,Y86}, the
variational method\cite{MC91,MRU95,GRD06} and perturbation theory\cite
{AILZ81,ACCG01,GRD06}. Regarding the latter approach we mention that the sum
over states is impractical for the calculation of corrections of order
greater than the first one\cite{AILZ81,ACCG01,GRD06}. It is preferable
either to integrate the perturbation equations directly\cite{F01} or to make
use of the hypervirial and Hellmann--Feynman theorems\cite{F01,FC87}.

In the case of the harmonic oscillator, most studies refer to the case of a
particle that moves in a box under the effect of a potential of the form $%
V(x)=k(x-x_{0})^{2}/2$ as if it were tied to the point $x_{0}$ by means of a
spring of force constant $k$\cite{MC88,MC91,RRML00,ACCG01,GRD06}. If we
assume that the Hooke's force is due to the interaction between two
particles then we have a model similar to the one discussed by Tanner\cite
{T91} who showed that it is not possible to separate the center of mass and
internal degrees of freedom in the usual way because of the effect of the
boundary conditions. Amore and Fern\'{a}ndez\cite{AF09} have recently
discussed this problem in greater detail.

The usual model for the confined hydrogen atom suffers from the same
limitation: the nucleus is considered to be clamped somewhere inside the box%
\cite{Y86,W89,B91,MC91,LM93,MRU95}. It appears to be most interesting to
assume that not only the electron but also the nucleus moves inside it. The
purpose of this paper is to discuss such a model in the simplest possible
ways. In Sec.~\ref{sec:model} we outline the model and write the Schr\"{o}%
dinger equation in a dimensionless form. In Sec.~\ref{sec:PT} we obtain the
ground--state energy approximately by means of first--order perturbation
theory and compare it with the one for the clamped--nucleus case. In Sec.~%
\ref{sec:Variational} we resort to a somewhat more elaborate variational
function. Finally, in Sec.~\ref{sec:conclusions} we summarize and discuss
the results.

\section{The model}

\label{sec:model}

The Hamiltonian operator for a nonrelativistic hydrogen--like atom is
\begin{eqnarray}
\hat{H} &=&\hat{T}+\hat{V}  \nonumber \\
\hat{T} &=&-\frac{\hbar ^{2}}{2m_{e}}\nabla _{e}^{2}-\frac{\hbar ^{2}}{2m_{n}%
}\nabla _{n}^{2}  \nonumber \\
V(r) &=&-\frac{Ze^{2}}{4\pi \epsilon _{0}r}  \label{eq:H}
\end{eqnarray}
where $m_{e}$ and $m_{n}$ are the masses of the electron and nucleus located
at $\mathbf{r}_{e}$ and $\mathbf{r}_{n}$ with charges $-e$ and $Ze$,
respectively, $r=|\mathbf{r|}$, $\mathbf{r}=\mathbf{r}_{e}-\mathbf{r}_{n}$, $%
\epsilon _{0}$ is the vacuum permittivity and $\nabla ^{2}$ denotes the
Laplacian in the coordinates indicated by the subscript.

In the case of the free atom we separate the motion of the center of mass
from the internal one by means of a well--known change of variables and
obtain
\begin{eqnarray}
\hat{H} &=&\hat{H}_{CM}+\hat{H}_{int}  \nonumber \\
\hat{H}_{CM} &=&-\frac{\hbar ^{2}}{2M}\nabla _{CM}^{2},\;M=m_{e}+m_{n}
\nonumber \\
\hat{H}_{int} &=&-\frac{\hbar ^{2}}{2m}\nabla ^{2}+V(r),\;m=\frac{m_{e}m_{n}%
}{M}  \label{eq:H_CM}
\end{eqnarray}
where $\nabla ^{2}$ and $\nabla _{CM}^{2}$ are the Laplacians for the
variables $\mathbf{r}$ and $\mathbf{r}_{CM}=(m_{e}\mathbf{r}_{e}+m_{n}%
\mathbf{r}_{n})/M$, respectively, and $m$ is the reduced mass. Thus, we can
factor the energy states of the free hydrogen atom as $\psi (\mathbf{r}_{e},%
\mathbf{r}_{n})=\psi _{CM}(\mathbf{r}_{CM})\psi _{int}(\mathbf{r})$ and
solve the Schr\"{o}dinger equation for $\hat{H}_{int}$ in the usual way\cite
{EWK44,P68}. The eigenfunctions and eigenvalues of $\hat{H}_{int}$ provide
all the physical properties of the isolated atom, such as, for example, the
spectral lines, selection rules, etc.\cite{EWK44,P68}.

If the atom is confined to a spherical box of radius $R$ with impenetrable
walls then the states should vanish when either $r_{e}=R$ or $r_{n}=R$ and,
consequently, the above separation is not possible as discussed by Tanner%
\cite{T91} and Amore and Fern\'{a}ndez\cite{AF09} for the harmonic
oscillator. The reason is that the variables $\mathbf{r}_{CM}$ and $\mathbf{r%
}$ are unsuitable for the boundary conditions that are naturally given in
terms of $\mathbf{r}_{e}$ and $\mathbf{r}_{n}$.

The positions of the electron an nucleus in the box are completely
determined by six variables. We conveniently choose $r_{e}$, $r_{n}$ and $r$
(the sides of a triangle) plus three angles for the orientation of the
triangle in space. The $S$ states (those with zero angular momentum) depend
only on the three radial variables: $\psi (r_{e},r_{n},r\dot{)}$. In fact,
if we take into account that
\begin{eqnarray}
\nabla _{e}\psi  &=&\frac{\mathbf{r}_{e}}{r_{e}}\frac{\partial \psi }{%
\partial r_{e}}+\frac{\mathbf{r}}{r}\frac{\partial \psi }{\partial r}
\nonumber \\
\nabla _{n}\psi  &=&\frac{\mathbf{r}_{n}}{r_{n}}\frac{\partial \psi }{%
\partial r_{n}}-\frac{\mathbf{r}}{r}\frac{\partial \psi }{\partial r}
\end{eqnarray}
then we realize that $\psi (r_{e},r_{n},r\dot{)}$ has zero angular momentum:
\begin{equation}
\left( \mathbf{r}_{e}\times \nabla _{e}+\mathbf{r}_{n}\times \nabla
_{n}\right) \psi =\frac{(\mathbf{r}_{e}-\mathbf{r}_{n})\times \mathbf{r}}{r}%
\frac{\partial \psi }{\partial r}=0
\end{equation}

In order to simplify the calculation we first make the change of variables $%
\mathbf{r}_{e}^{\prime }=\mathbf{r}_{e}/R$ and $\mathbf{r}_{n}^{\prime }=%
\mathbf{r}_{n}/R$ that leads to the dimensionless Hamiltonian operator
\begin{eqnarray}
\hat{H}_{d} &=&\frac{m_{e}R^{2}}{\hbar ^{2}}\hat{H}=-\frac{1}{2}\nabla
_{e}^{\prime 2}-\frac{\beta }{2}\nabla _{n}^{\prime 2}-\frac{\lambda }{%
r^{\prime }}  \nonumber \\
\beta &=&\frac{m_{e}}{m_{n}},\;\lambda =\frac{m_{e}Ze^{2}R}{4\pi \epsilon
_{0}\hbar ^{2}}  \label{eq:Hd}
\end{eqnarray}
The states of this dimensionless system vanish when either $\mathbf{r}%
_{e}^{\prime }=1$ or $\mathbf{r}_{n}^{\prime }=1$. From now on we omit the
primes on the dimensionless quantities but keep in mind that lengths, masses
and energies are measured in units of $R$, $m_{e}$, and $\hbar
^{2}/(m_{e}R^{2})$, respectively. For example, $1/\beta $ is the nuclear
mass in such units.

\section{Perturbation theory}

\label{sec:PT}

For simplicity we restrict ourselves to the ground state and a small box
radius. If $\lambda $ is a small parameter then we can try perturbation
theory in terms of the unperturbed or reference Hamiltonian $\hat{H}_{d}^{0}=%
\hat{H}_{d}(\lambda =0)$. The perturbation is therefore given by the
interaction between the particles $\hat{H}_{d}^{\prime }=-1/r$. The
unperturbed ground state is
\begin{equation}
\varphi (r_{e},r_{n})=2\frac{\sin (\pi r_{e})\sin (\pi r_{n})}{r_{e}r_{n}}
\label{eq:phi_0}
\end{equation}
Therefore, the expectation value of $\hat{H}_{d}$ with this function gives
us the energy of the ground state corrected through first order of
perturbation theory. Besides, according to the variation principle such
approximate energy will be an upper bound to the exact one\cite{EWK44,P68}.

The calculation is reminiscent of that for the helium atom under the
clamped--nucleus approximation and we may therefore profit from well--known
results. The calculation of the expectation value of the kinetic energy is
straightforward, and there are various ways of calculating the expectation
value of $1/r$\cite{EWK44,P68}. Here, we resort to the expansion of $1/r$ in
terms of Legendre polynomials that leads to the simple integral\cite{EWK44}
\begin{eqnarray}
\int \frac{\varphi (r_{e},r_{n})^{2}}{r}d\tau _{e}d\tau _{n} &=&16\pi
^{2}\left[ \int_{0}^{1}\int_{0}^{r_{e}}\varphi
(r_{e},r_{n})^{2}r_{e}r_{n}^{2}dr_{n}dr_{e}\right.  \nonumber \\
&&\left. +\int_{0}^{1}\int_{r_{e}}^{1}\varphi
(r_{e},r_{n})^{2}r_{e}^{2}r_{n}dr_{n}dr_{e}\right]  \label{eq:<1/r>}
\end{eqnarray}
Since the analytical expression is rather cumbersome we just show the
numerical result:
\begin{eqnarray}
\epsilon (\lambda ) &=&\frac{\pi ^{2}(\beta +1)}{2}-1.786073167\lambda
\nonumber \\
&=&4.934802200(\beta +1)-1.786073167\lambda  \label{eq:e(lam)_PT}
\end{eqnarray}

Computer algebra systems are nowadays available in the science departments
of most universities because they are invaluable teaching tools. This
problem may be useful for motivating the students to resort to such software.

We can obtain simple analytical expressions by means of the even simpler
trial function
\begin{equation}
\varphi (r_{e},r_{n})=30(1-r_{e})(1-r_{n})  \label{eq:phi_0_b}
\end{equation}
that leads to a quite similar result
\begin{eqnarray}
\epsilon (\lambda ) &=&5(\beta +1)-\frac{25\lambda }{14}  \nonumber \\
&=&5(\beta +1)-1.785714285\lambda  \label{eq:e(lam)_anal}
\end{eqnarray}

\begin{figure}[H]
\begin{center}
\includegraphics[width=9cm]{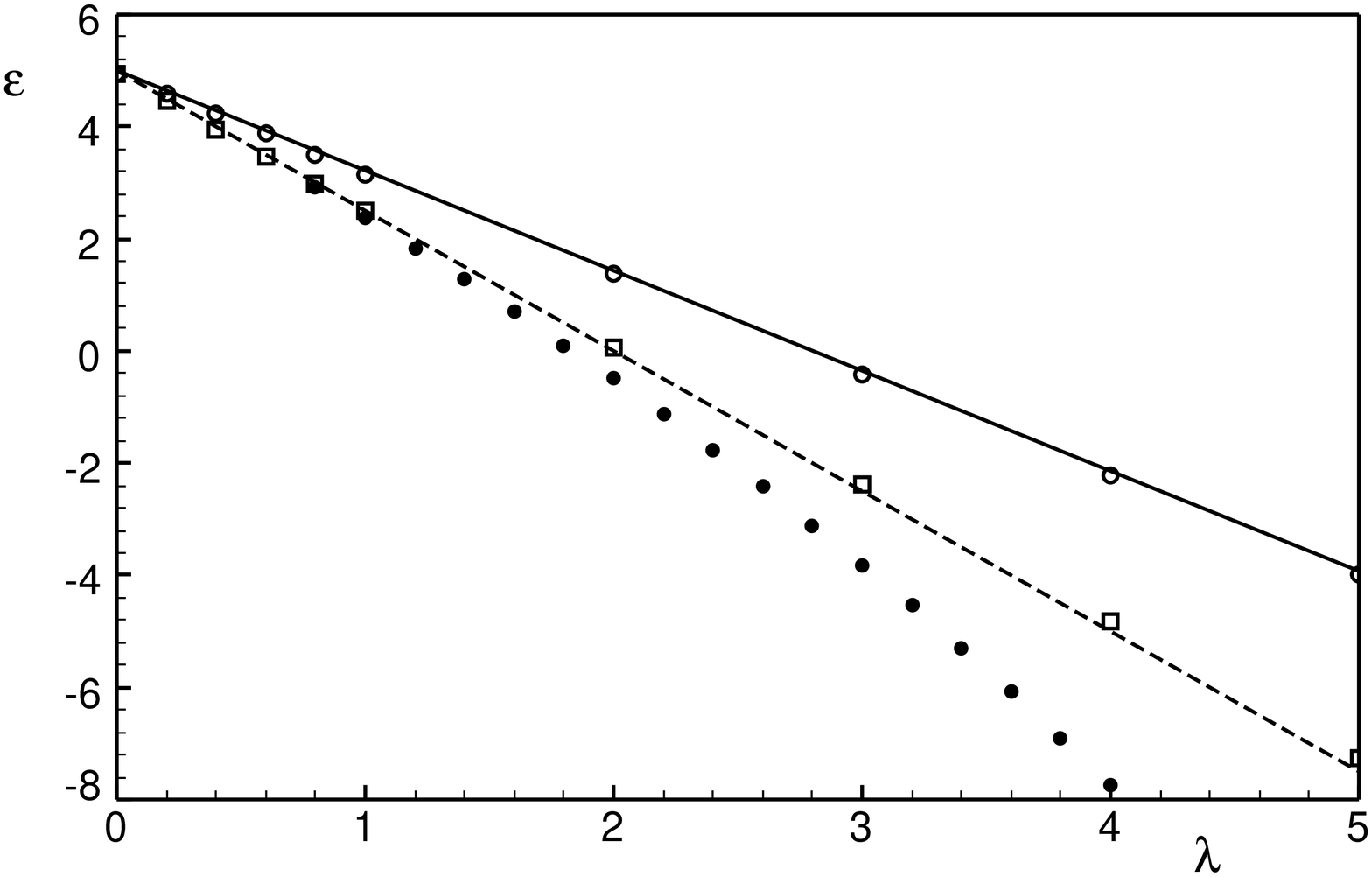}
\end{center}
\caption{Results from equations (\ref{eq:e(lam)_PT}) (circles), (\ref
{eq:e(lam)_anal}) (solid line), (\ref{eq:eH(lam)_anal}) (dashed line), (\ref
{eq:eH(lam)_PT}) (squares) and power series (filled circles).}
\label{fig:BH}
\end{figure}

It is interesting to compare the results for this model with those for the
hydrogen atom with the nucleus clamped at the center of the box. If we
calculate the expectation value of the dimensionless Hamiltonian operator
(notice that we use the same units as before)
\begin{equation}
\hat{H}_{dH}=-\frac{1}{2}\nabla ^{2}-\frac{\lambda }{r}  \label{eq:HdH}
\end{equation}
with the approximate trial function $\varphi (r)=\sqrt{30}(1-r)$ we obtain
\begin{equation}
\epsilon _{H}(\lambda )=5-\frac{5\lambda }{2}  \label{eq:eH(lam)_anal}
\end{equation}
For comparison we also consider the unperturbed ground state
\begin{equation}
\varphi (r)=\sqrt{2}\frac{\sin (\pi r)}{r}  \label{eq:phi_H_0}
\end{equation}
that leads to the first--order dimensionless energy
\begin{eqnarray}
\epsilon _{H}(\lambda ) &=&\frac{\pi ^{2}}{2}-2.437653392\lambda   \nonumber
\\
&=&4.934802200-2.437653392\lambda   \label{eq:eH(lam)_PT}
\end{eqnarray}
Obviously, these results are valid for sufficiently small values of $\lambda
$. After contrasting equations (\ref{eq:eH(lam)_anal}) and (\ref
{eq:eH(lam)_PT}) with more accurate results\cite{F01} we conclude that
present first--order estimates are acceptable for $\lambda \lesssim 1$. In
principle, we may assume that the accuracy of present moving--nucleus
results are as accurate as the clamped--nucleus ones for $\lambda \lesssim 1$
. If this is true, then our results suggest that the energy of the
moving--nucleus model is larger than the clamped--nucleus one ($\epsilon
(\lambda )>\epsilon _{H}(\lambda )$), at least for sufficiently small box
radii. The difference does not come mainly from the kinetic energy of the
nucleus that is proportional to $\beta \approx 1/1836$ but from the
electron--nucleus interaction. This conclusion is consistent with earlier
variational results that show that the smallest energy takes place when the
nucleus is clamped at the center of the sphere and increases as it
approaches the wall\cite{MRU95}. Fig.~\ref{fig:BH} shows the approximate
energies given by equations (\ref{eq:e(lam)_PT}), (\ref{eq:e(lam)_anal}), (%
\ref{eq:eH(lam)_anal}) and (\ref{eq:eH(lam)_PT}) for $\lambda \leq 5$ as
well as accurate numerical energies for the clamped nucleus model calculated
by a straightforward power--series method\cite{F01}.

The critical value of $\lambda $ defined by $\epsilon _{H}(\lambda _{c})=0$
estimated from the first--order perturbation energy (\ref{eq:eH(lam)_PT}) $%
\lambda _{c}\approx 2$ is about $9\%$ larger than the actual value $\lambda
_{c}=1.835246330$ that one easily obtains by means of the method already
mentioned above or from perturbation theory of greater order\cite{F01}. For
the moving--nucleus model our approximate expressions (\ref{eq:e(lam)_PT})
and (\ref{eq:e(lam)_anal}) predict $\lambda _{c}\approx 2.8$ and we expect
that its error is of comparable magnitude.

\section{Variational method}

\label{sec:Variational}

We can improve the results of the preceding section by means of the
variational method. To this end we propose the trial function
\begin{equation}
\varphi (r_{e},r_{n},r)=N(\alpha )(1-r_{e})(1-r_{n})e^{-\alpha r}
\label{eq:phi_v}
\end{equation}
where $\alpha $ is a variational parameter and $N(\alpha )$ an appropriate
normalization factor. Since it has the correct asymptotic behaviour $%
e^{-\alpha r}$ for the free atom, we expect it to yield accurate energies
for large values of $\lambda $. More precisely, we expect that it yields the
exact energy of the ground state of the free hydrogen
\begin{equation}
\lim_{\lambda \rightarrow \infty }\frac{\epsilon }{\lambda ^{2}}=-\frac{1}{%
2(1+\beta )}\approx -\frac{1}{2}  \label{eq:lim_epsilon}
\end{equation}

By means of a straightforward but tedious calculation we prove that
\begin{eqnarray}
\hat{H}_{d}\varphi  &=&-\frac{1}{2}\left( \frac{\partial ^{2}}{\partial
r_{e}^{2}}+\frac{2}{r_{e}}\frac{\partial }{\partial r_{e}}+\frac{%
r_{e}^{2}-r_{n}^{2}+r^{2}}{r_{e}r}\frac{\partial ^{2}}{\partial
r_{e}\partial r}\right) \varphi   \nonumber \\
&&-\frac{\beta }{2}\left( \frac{\partial ^{2}}{\partial r_{n}^{2}}+\frac{2}{%
r_{n}}\frac{\partial }{\partial r_{n}}+\frac{r_{n}^{2}-r_{e}^{2}+r^{2}}{%
r_{n}r}\frac{\partial ^{2}}{\partial r_{n}\partial r}\right) \varphi
\nonumber \\
&&-\frac{1+\beta }{2}\left( \frac{\partial ^{2}}{\partial r^{2}}+\frac{2}{r}%
\frac{\partial }{\partial r}\right) \varphi -\frac{\lambda }{r}\varphi
\label{eq:Hd_phi_v}
\end{eqnarray}

It follows from $r^{2}=r_{e}^{2}+r_{n}^{2}-2\mathbf{r}_{e}\cdot \mathbf{r}%
_{n}$ and $-r_{e}r_{n}\leq \mathbf{r}_{e}\cdot \mathbf{r}_{n}\leq r_{e}r_{n}$
that $|r_{e}-r_{n}|\leq r\leq r_{e}+r_{n}$. Therefore, we can calculate all
the integrals that are necessary for the variational approach by means of
the simple recipe\cite{H64}
\begin{eqnarray}
\int \int \int f(r_{e},r_{n},r)\,d\tau
&=&\int_{0}^{1}\int_{0}^{r_{e}}%
\int_{r_{e}-r_{n}}^{r_{e}+r_{n}}f(r_{e},r_{n},r)r_{e}r_{n}r\,dr_{e}\,dr_{n}%
\,dr  \nonumber \\
&&+\int_{0}^{1}\int_{0}^{r_{n}}%
\int_{r_{n}-r_{e}}^{r_{e}+r_{n}}f(r_{e},r_{n},r)r_{e}r_{n}r\,dr_{n}\,dr_{e}%
\,dr
\end{eqnarray}

The calculation of the variational energy is simple but tedious and yields
rather cumbersome results. In what follows we simply outline the main steps.
First, notice that
\begin{equation}
\epsilon (\alpha ,\lambda )=\left\langle \hat{H}_{d}\right\rangle
=\left\langle \hat{T}\right\rangle (\alpha )-\lambda \left\langle
1/r\right\rangle (\alpha )
\end{equation}
is a linear function of $\lambda $. Therefore, instead of calculating $%
\alpha (\lambda )$ from the minimum condition $\partial \epsilon (\alpha
,\lambda )/\partial \alpha =0$ we derive an analytical expression for $%
\lambda (\alpha )$ and a parametric expression for the variational energy: $%
\epsilon (\alpha ,\lambda (\alpha ))$. One can easily carry out this
calculation by means of a computer algebra system.

\begin{figure}[H]
\begin{center}
\includegraphics[width=9cm]{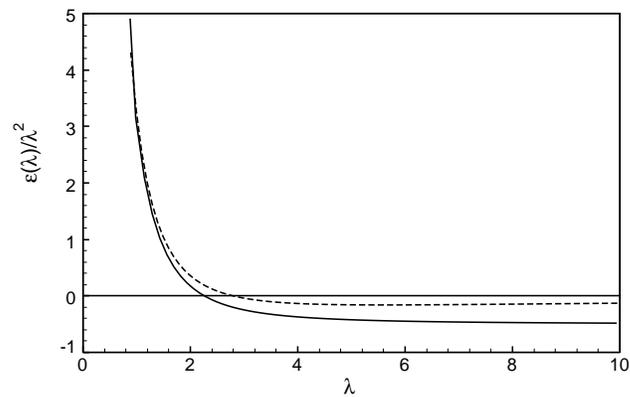}
\end{center}
\caption{Approximate values of $\frac{\epsilon(\lambda)}{\lambda^2}$
obtained from the trial functions (\ref{eq:phi_0_b}) (dashed line) and
(\ref{eq:phi_v}) (solid line).}
\label{fig:BHV}
\end{figure}

Fig.~\ref{fig:BHV} shows approximate values of $\epsilon /\lambda ^{2}$
obtained from the trial functions (\ref{eq:phi_0_b}) and (\ref{eq:phi_v}).
We appreciate that there is a satisfactory agreement for $\lambda \lesssim 1$
as expected but it seems that our earlier prediction of the critical value $%
\lambda _{c}$ was not as accurate as we believed. The variational function (%
\ref{eq:phi_v}) yields the more accurate value $\lambda _{c}=2.262$ that
differs $21\%$ from our earlier estimation. We also realize that the
variational energy approaches the exact free--atom energy (\ref
{eq:lim_epsilon}) as $\lambda $ increases.

In passing, we mention that the variational calculation confirms the earlier
perturbation result that the energy of the moving--nucleus atom is larger
than the one with the nucleus clamped at origin.

\section{Conclusions}

\label{sec:conclusions}

Tanner\cite{T91} proposed a pedagogical discussion of the effect of the
boundary conditions on the separability of the degrees of freedom of a
confined system. However, he did not show any result for the
one--dimensional harmonic oscillator that he chose as an illustrative
example. Later Amore and Fern\'{a}ndez\cite{AF09} discussed that model in
more detail. In this article we extended those arguments to the hydrogen
atom and carried out simple approximate calculations for the ground state by
means of straightforward first--order perturbation theory and a more
accurate variational approach. It has been our purpose to show how to do the
calculation using well--known techniques already applied to the helium atom
with the clamped--nucleus approximation. Our analysis shows that it is not
possible to separate the Schr\"{o}dinger equation in the usual way in terms
of internal and center--of--mass coordinates $\mathbf{r}$ and $\mathbf{r}%
_{CM}$, respectively, because the boundary conditions are given in terms of
the electron and nucleus coordinates $\mathbf{r}_{e}$, and $\mathbf{r}_{n}$,
respectively. Our approximate results show that the energy is greater when
the nucleus moves than when it is clamped at the center of the spherical box
and that the difference does not come mainly from the kinetic energy of the
moving nucleus that is considerably smaller that the electronic one. Present
results based on perturbation theory are limited to a small box radius or
strong confinement but the variational ones are valid for all box sizes and
yield the free--atom energy when $R\rightarrow \infty $. Although it is
relatively easy to carry out perturbation calculations of large order for
the clamped--nucleus model\cite{F01}, the treatment of the moving--nucleus
case is considerably more complicated. The variational method appears to be
a better choice. It can be improved as in the case of the Helium atom by
means of a Hylleraas--like expansion\cite{H64} for the trial function $%
\varphi (r_{e},r_{n},r)$.

\end{document}